\documentclass[pra,twocolumn,amsmath]{revtex4}

\usepackage{hyperref}
\usepackage{graphicx}
\usepackage{bm}
\usepackage{natbib}
\usepackage{amsmath}
\usepackage{braket}
\usepackage{siunitx}
\usepackage{dcolumn}
\usepackage{color}
\usepackage[normalem]{ulem} 

\newcommand{\state}[3]{{}^{#1}\mathrm{#2}_{#3}}
\newcommand{\isotope}[2]{{}^{#1}\mathrm{#2}}
\newcommand{\Sr}{\isotope{87}{Sr}}
\newcommand{\kB}{k_\mathrm{B}}							
\newcommand{\Erec}{E_\mathrm{r}} 						
\newcommand{\TrapFreqAx}{\nu_z}							
\newcommand{\Trad}{T_r}									
\newcommand{\VEFraction}{\epsilon}						
\newcommand{\Ngnd}{N_\mathrm{g}}						
\newcommand{\Nexc}{N_\mathrm{e}}						
\newcommand{\Nany}{N_\mathrm{e,g}}						
\newcommand{\thold}{t_\mathrm{h}}						
\newcommand{\Upeak}{U_0}								
\newcommand{\Utherm}{U}									
\newcommand{\traplossGS}{\Gamma_{\mathrm{bg}}}			
\newcommand{\traplossES}{\Gamma_{\mathrm{bg}}^\prime}	
\newcommand{\traplossDIFF}{\Delta\Gamma_{\mathrm{bg}}}	
\newcommand{\decayrateLAT}{\Gamma_{\mathrm{L}}}			
\newcommand{\decayratecoeffLAT}{\gamma_{\mathrm{L}}}	
\newcommand{\decayrateOTHER}{\Gamma_{\mathrm{0}}}		
\newcommand{\decayrateBBR}{\Gamma_{\mathrm{BBR}}}		
\newcommand{\decayrateSPO}{\Gamma_{\mathrm{s}}}			
\newcommand{\branchingratio}{R}							
\newcommand{\threejsymbol}[6]{\left( \begin{array}{ccc} #1 & #2	& #3 \\ #4 & #5	& #6 \end{array} \right)}
\newcommand{\sixjsymbol}[6]{\left\{ \begin{array}{ccc} #1 & #2	& #3 \\ #4 & #5	& #6 \end{array} \right\}}
\newcommand{\matrixelement}[3]{\braket{#1|#2|#3}}
\newcommand{\reducedmatrixelement}[3]{\braket{#1||#2||#3}}

\begin{document}
\preprint{}
\title{Lattice-induced photon scattering in an optical lattice clock}
\author{S\"oren D\"orscher}
\author{Roman Schwarz}
\author{Ali Al-Masoudi}
\author{Stephan Falke}\altaffiliation{Present address: TOPTICA Photonics AG, Lochhamer Schlag 19, 82166 Gr\"afelfing, Germany}
\author{Uwe Sterr}
\author{Christian Lisdat}\email{christian.lisdat@ptb.de}
\affiliation{Physikalisch-Technische Bundesanstalt, Bundesallee 100, 38116 Braunschweig, Germany}
\date{\today}
\begin{abstract}
We investigate scattering of lattice laser radiation in a strontium optical lattice clock and its implications for operating clocks at interrogation times up to several tens of seconds.
Rayleigh scattering does not cause significant decoherence of the atomic superposition state near a magic wavelength.
Among the Raman scattering processes, lattice-induced decay of the excited state $(5s5p)\,\state{3}{P}{0}$ to the ground state $(5s^2)\,\state{1}{S}{0}$ via the state $(5s5p)\,\state{3}{P}{1}$ is particularly relevant, as it reduces the effective lifetime of the excited state and gives rise to quantum projection noise in spectroscopy.
We observe this process in our experiment and find a decay rate of $\SI{556(15)d-6}{\per\second}$ per photon recoil energy $\Erec$ of effective lattice depth, which agrees well with the rate we predict from atomic data.
We also derive a natural lifetime $\tau = \SI{330(140)}{\second}$ of the excited state $\state{3}{P}{0}$ from our observations.
Lattice-induced decay thus exceeds spontaneous decay at typical lattice depths used by present clocks.
It eventually limits interrogation times in clocks restricted to high-intensity lattices, but can be largely avoided, e.g., by operating them with shallow lattice potentials.
\end{abstract}
%
\maketitle
\section{\label{sec:intro}Introduction}
Atomic clocks based on optical transitions keep advancing the field of frequency metrology.
Accuracy of few parts in $10^{18}$ \cite{ush15, nic15, nem16, hun16} as well as fractional frequency instability below $10^{-16} / \sqrt{\tau}$ \cite{sch17} have been demonstrated, where $\tau$ is the integration time in seconds.
Present optical clocks are based either on single trapped ions or on large numbers of neutral atoms confined to optical lattice potentials, as has first been proposed in Ref.~\cite{kat02}.
Both types of clocks lend themselves to a variety of new applications, which range from testing fundamental physics with laboratory experiments, such as searching for variations of fundamental constants \cite{hun14,god14}, testing special relativity \cite{del17}, or the proposed search for dark matter \cite{der14}, to the measurement of geopotentials \cite{tak16,lis16,gro18}.

The Allan deviation, $\sigma_y(\tau)$, which can be  achieved in a given integration time $\tau$ for an ensemble of $N$ uncorrelated absorbers is fundamentally limited by quantum projection noise (QPN) to \cite{rie04a}
\begin{equation}
	\sigma_y(\tau) = \frac{1}{K} \frac{1}{Q} \frac{1}{\sqrt{N}}
	\sqrt{\frac{T_\mathrm{c}}{\tau}} \mbox{,}
	\label{eq:QPNL}
\end{equation}
where $Q$ is the quality factor of the observed resonance, $T_\mathrm{c}$ is the measurement cycle duration, and $K$ is a line shape factor on the order of one, which depends on the interrogation sequence.
Optical lattice clocks benefit from inherently low QPN due to the large number of particles being probed simultaneously.
In fact, they are often limited by the Dick effect \cite{dic87,que03,alm15} instead, which results from the aliasing of noise due to the noncontinuous observation of the interrogation laser's frequency.
This effect can be overcome by dead time--free interrogation using two atomics packages \cite{sch17}, for instance.

Generally, the performance of optical clocks benefits strongly from increasing interrogation time and thus the quality factor.
Ramsey spectroscopy with free evolution times up to \SI{15}{\second} has been demonstrated in a three-dimensional optical lattice clock \cite{mar18}.
Ultrastable lasers achieving fractional frequency instability down to $4 \times 10^{-17}$ and coherence times of several tens of seconds have been reported recently \cite{mat17a}; such laser systems will allow even longer interrogation times.
However, new potential sources of atomic decoherence or frequency shifts will become relevant at these timescales. 

In this article, we investigate off-resonant scattering of lattice laser radiation by $\Sr$ atoms in the two states $\state{1}{S}{0}$ and $\state{3}{P}{0}$ and its effect on clock operation.
As in other strontium clocks, our one-dimensional lattice operates near the magic wavelength at $\SI{813}{\nano\meter}$, where the atomic polarizabilities of these two states are equal.
Typical potential depths are between $50\,\Erec$ and $200\,\Erec$ supporting four to nine longitudinally bound states, where $\Erec=h^2/(2m\lambda^2)$ is the photon recoil energy at the lattice wavelength $\lambda$ for an atom of mass $m$.
Some aspects of photon scattering have previously been studied theoretically \cite{mar13a}, but, to the best of our knowledge, these processes have not been investigated experimentally yet.
Potential effects on clock operation include decoherence of the atomic superposition state during interrogation, degradation of the signal-to-noise ratio and frequency stability, and systematic frequency shifts.
Especially for short excitation pulses, atoms interact near-resonantly with the interrogation laser after specific state-changing scattering events.
Observation of a resulting loss of contrast at long interrogation times has recently been reported for Ramsey spectroscopy \cite{mar18}.

\begin{figure}[tb]
	\centerline{\includegraphics[width=86 mm]{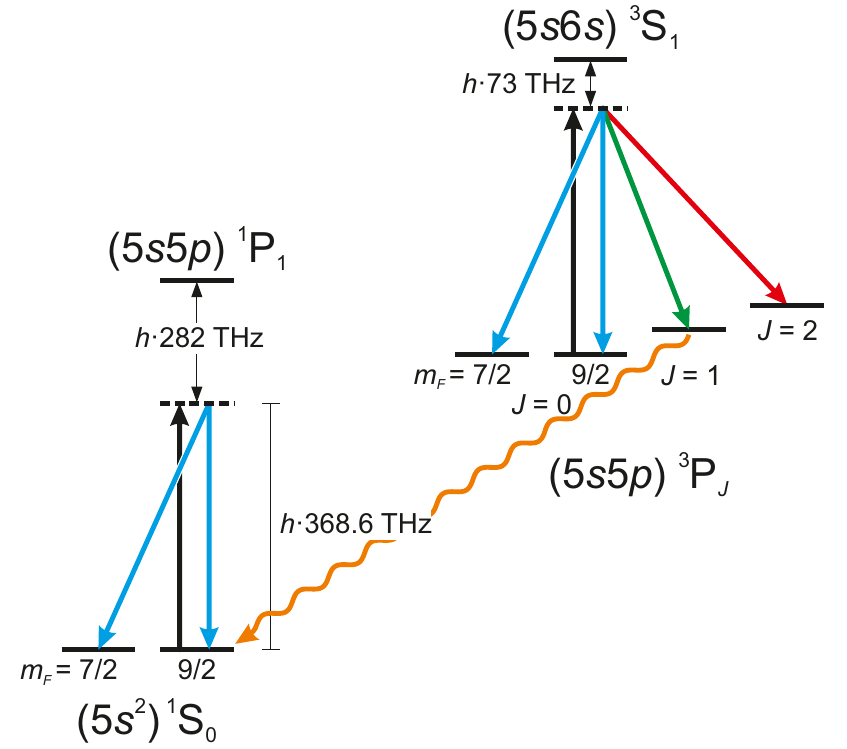}}
	\caption{(color online)
		Schematic summary of photon scattering near the magic wavelength $\lambda \approx \SI{813}{\nano\meter}$, i.e., a frequency of $\SI{368.6}{\tera\hertz}$, in the magnetic substates $m_F=9/2$ of the ground state and excited state of a $\Sr$ optical lattice clock.
		Only the dominant intermediate states are shown.}
	\label{fig:scattering_summary}
\end{figure}
Figure~\ref{fig:scattering_summary} schematically illustrates the different types of photon scattering processes at the lattice wavelength, all of which are off-resonant.
In the following, we refer to inelastic scattering events that change the internal state of the atom as Raman scattering and to the elastic ones that leave the atom's internal state unchanged as Rayleigh scattering.
As will be shown in Sec.~\ref{sec:scat}, the overall scattering rates into each of the final fine-structure states shown in Fig.~\ref{fig:scattering_summary} are similar in magnitude.
However, the resulting effects on an optical lattice clock can be quite different.

For instance, Raman scattering $\state{3}{P}{0} \rightarrow \state{3}{P}{1}$ is followed by radiative decay to the ground state, since the state $\state{3}{P}{1}$ has a lifetime of about \SI{21}{\micro\second} \cite{nic15}.
At typical potential depths $\Upeak \gg \Erec$, the atom is not lost from the trap in the process.
Thus, this off-resonant photon scattering process effectively induces decay of the excited state, in addition to other decay mechanisms such as spontaneous emission or pumping by blackbody radiation (BBR).
In contrast, the state $\state{3}{P}{2}$ has an effective lifetime on the order of \SI{100}{\second} at room temperature, which is limited by BBR-induced decay \cite{yas04}.
Raman scattering $\state{3}{P}{0} \rightarrow \state{3}{P}{2}$ thus leaves atoms shelved in the metastable state.

We study lattice-induced decay to the ground state $\state{1}{S}{0}$ experimentally in Sec.~\ref{sec:exp} and determine the rate of decay as a function of lattice depth.
Rayleigh scattering and Raman scattering into the state $\state{3}{P}{2}$ cannot be investigated experimentally with our setup, they are treated theoretically in Sec.~\ref{sec:scat} instead.
There, we present a complementary investigation of all off-resonant scattering processes, predicting their rates from atomic data.
To derive the natural lifetime of the excited state from our measurements, we estimate the decay rate of the excited state due to coupling to BBR in Sec.~\ref{sec:lifetime}.
The effects of photon scattering on atomic coherence and the performance of optical lattice clocks are analyzed in Sec.~\ref{sec:consequences}.
Finally, we discuss approaches to solve or avoid the potential problems at long interrogation times in Sec.~\ref{sec:discussion} and summarize our findings in Sec.~\ref{sec:conclusion}.
\section{\label{sec:exp}Lattice-induced decay of the excited state}
We investigate the decay rate $\state{3}{P}{0} \rightarrow \state{1}{S}{0}$ experimentally by measuring the populations in each of the two clock states as a function of hold time in the trap.
Lattice-induced scattering to the state $\state{3}{P}{1}$ is identified and discerned from other decay mechanisms by varying the depth of the optical lattice, since its rate is proportional to the intensity of the lattice laser field.
\subsection{\label{subsec:exp_setup}Experimental setup}
For our measurements, we operate our $\Sr$ lattice clock, which has been discussed in previous publications \cite{gre16, alm15, fal14}, as follows.

After two-stage laser cooling, several hundred atoms at a temperature $T \approx \SI{2}{\micro\kelvin}$ are trapped in the optical lattice at an initial trap depth of about $100\,\Erec$.
Population of the stretched magnetic substate $m_F = 9/2$ is enhanced by optical pumping, using a laser beam which is resonant with the $F=F^\prime=9/2$ hyperfine component of the intercombination transition at \SI{689}{\nano\meter}, in a low bias magnetic field.

In order to prepare the atoms in the lowest two axial vibrational states of the lattice, we reduce its depth to about $50\,\Erec$ for \SI{40}{\milli\second}.
Atoms with axial vibrational quantum numbers $n_z > 1$ escape from the trap during this time, as these motional states are no longer trapped.
Subsequently, lattice depth is increased to its final value.

To prepare a pure atomic sample in the magnetic substate $m_F = 9/2$ of the excited state, we apply a strong bias magnetic field of about $\SI{0.6}{\milli\tesla}$, which splits the $\pi$-transitions of adjacent magnetic sublevels by approximately \SI{0.7}{\kilo\hertz}, and selectively transfer atoms from the magnetic substate $m_F = 9/2$ of the ground state to the excited state by a resonant $\pi$-pulse of the interrogation laser beam with a duration of about \SI{35}{\milli\second}.
The remaining ground-state population is removed from the trap by irradiating a laser beam on the strong transition $\state{1}{S}{0} \textendash \state{1}{P}{1}$.

After preparation, the sample is held in the optical lattice for a variable amount of time $\thold$.

Finally, we destructively detect the populations in the ground state, $\state{1}{S}{0}$, and in the metastable states, $\state{3}{P}{0}$ and $\state{3}{P}{2}$. As described previously \cite{fal14, alm15}, they are measured by collecting laser-induced fluorescence on the cycling transition $\state{1}{S}{0} \textendash \state{1}{P}{1}$ on a photomultiplier tube, and atoms in the metastable states are optically pumped to the ground state for detection.

For each hold time $\thold$ and lattice potential depth $\Upeak$, the populations $\Ngnd$ in $\state{1}{S}{0}$ and $\Nexc$ in $\state{3}{P}{0,2}$ are averaged over typically thirty to forty samples.
In order to reject long-term fluctuations of the initial number of atoms, they are also divided by the total population found in reference measurements with $\thold = \SI{1}{\second}$, which are performed before and after each set of measurements, for normalization.

Our detection scheme can neither resolve any magnetic substates nor distinguish the metastable states $\state{3}{P}{0}$ and $\state{3}{P}{2}$, because the atoms are repumped via the intermediate state $\state{3}{P}{1}$ by driving the transitions $\state{3}{P}{0,2} \textendash \state{3}{S}{1}$.
Therefore, only lattice-induced decay to the ground state (see Fig.~\ref{fig:scattering_summary}) can be studied experimentally in our system, whereas the lattice-induced population redistribution over the Zeeman and hyperfine states of the metastable $\state{3}{P}{}$ states cannot be resolved.
\subsection{\label{subsec:lattice_parameters}Optical lattice}
Laser light near the magic wavelength of $\Sr$ at \SI{813}{\nano\meter} is generated by a Ti:sapphire laser.
The optical lattice is formed by a single horizontal laser beam that is focused to a $1/e^2$-radius of about \SI{65}{\micro\meter} and retroreflected.

Sideband spectra of the reference transition $\state{1}{S}{0} \textendash \state{3}{P}{0}$ are used to determine the axial trapping frequencies $\TrapFreqAx$, fractions $\VEFraction$ of atoms in excited axial vibrational states, and radial temperatures $\Trad$ at each lattice depth.

The lattice potential depths $\Upeak$ are determined from $\TrapFreqAx$.
However, the effective lattice depth $\Utherm$ experienced by the atoms is smaller than $\Upeak$ due to the finite potential energy arising from atomic motion.
We account for the eigenenergy of the axial state with quantum number $n_z = 0$ in harmonic single-site approximation and neglect atoms in excited axial vibrational states ($n_z > 0$), which amount to a small fraction $\VEFraction \approx 0.05$ only.
For the radial degrees of freedom, we approximate the thermally averaged energy by its classical value, which is well justified for radial trapping frequencies of few \SI{100}{\hertz} and radial temperatures $\Trad \approx \SI{2}{\micro\kelvin}$.
Using the virial theorem, we find 
\begin{equation}
	\Utherm \approx \Upeak - \frac{1}{2} \left( \frac{1}{2} h \TrapFreqAx + 2 \kB \Trad \right) \mbox{.}
	\label{eq:thermal_lattice_depth}
\end{equation}
Most importantly, the lattice depth $\Utherm$ is proportional to the average light intensity experienced by the atoms.
We refer to it simply as the lattice depth in the following, unless stated otherwise.

For our measurements, we operate the lattice at potential depths $\Upeak$ between $52\,\Erec$ and $171\,\Erec$, which correspond to effective lattice depths $\Utherm$ between $37\,\Erec$ and $151\,\Erec$.
\subsection{\label{subsec:exp_results}Results}
\begin{figure}[tb]
	\centerline{\includegraphics[width=86 mm]{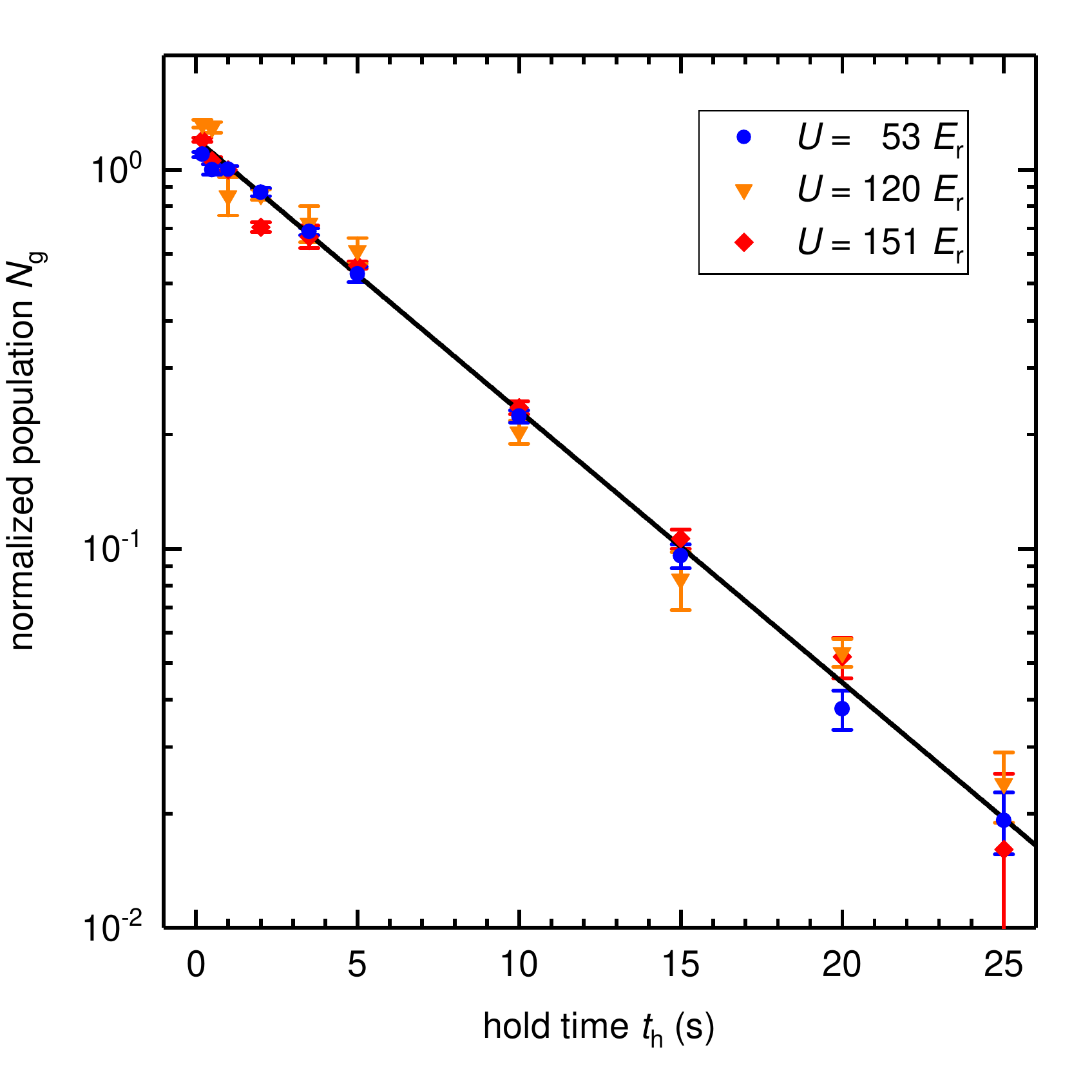}}
	\caption{(color online)
		Measured populations of samples prepared in the ground state at different effective lattice depths $\Utherm$ as a function of hold time $\thold$.
		The result of a combined fit of a rate equation model (see text for details) is shown as a solid line.}
	\label{fig:measurement_gs}
\end{figure}
\begin{figure*}[htb]
	\centerline{\includegraphics[width=86 mm]{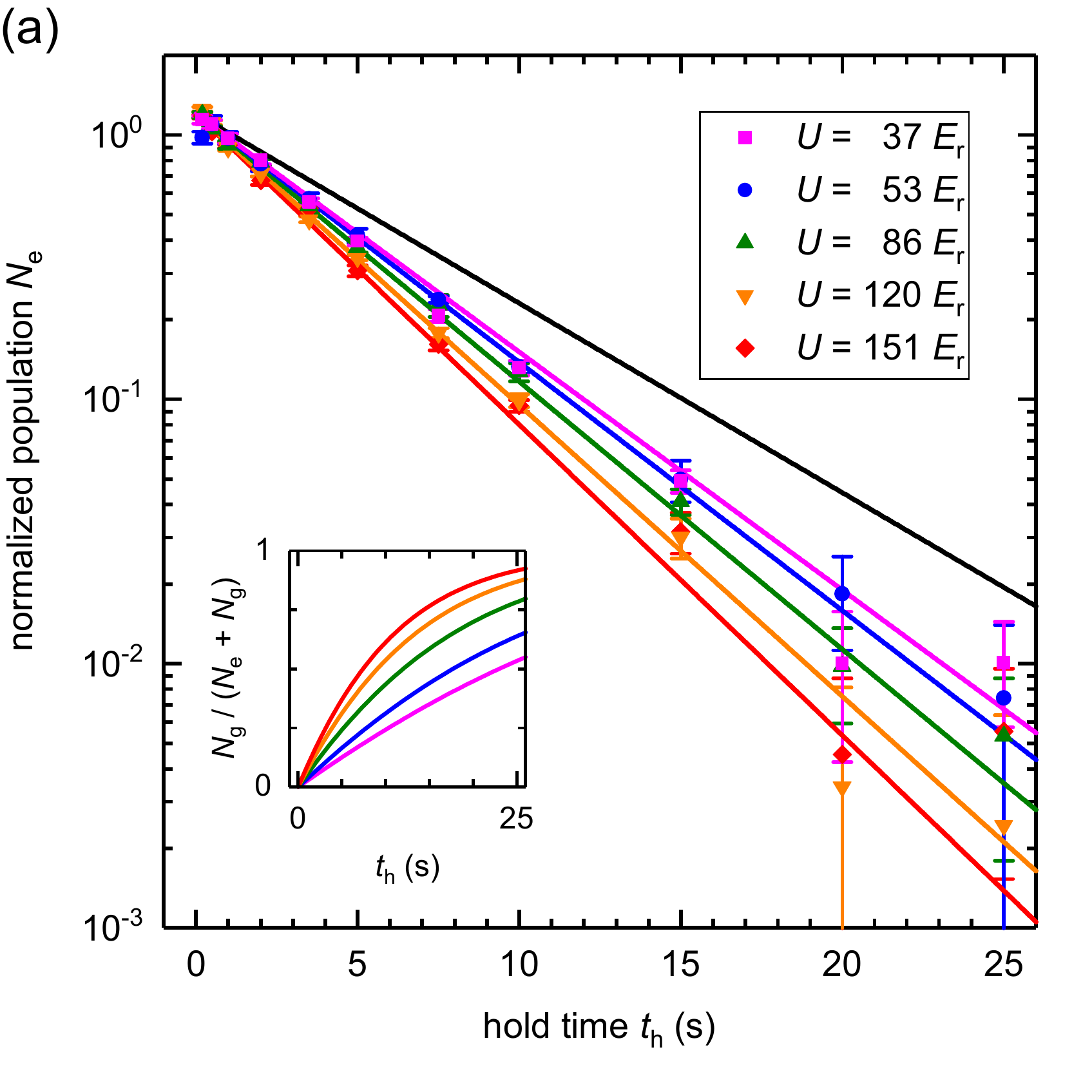}
		\includegraphics[width=86 mm]{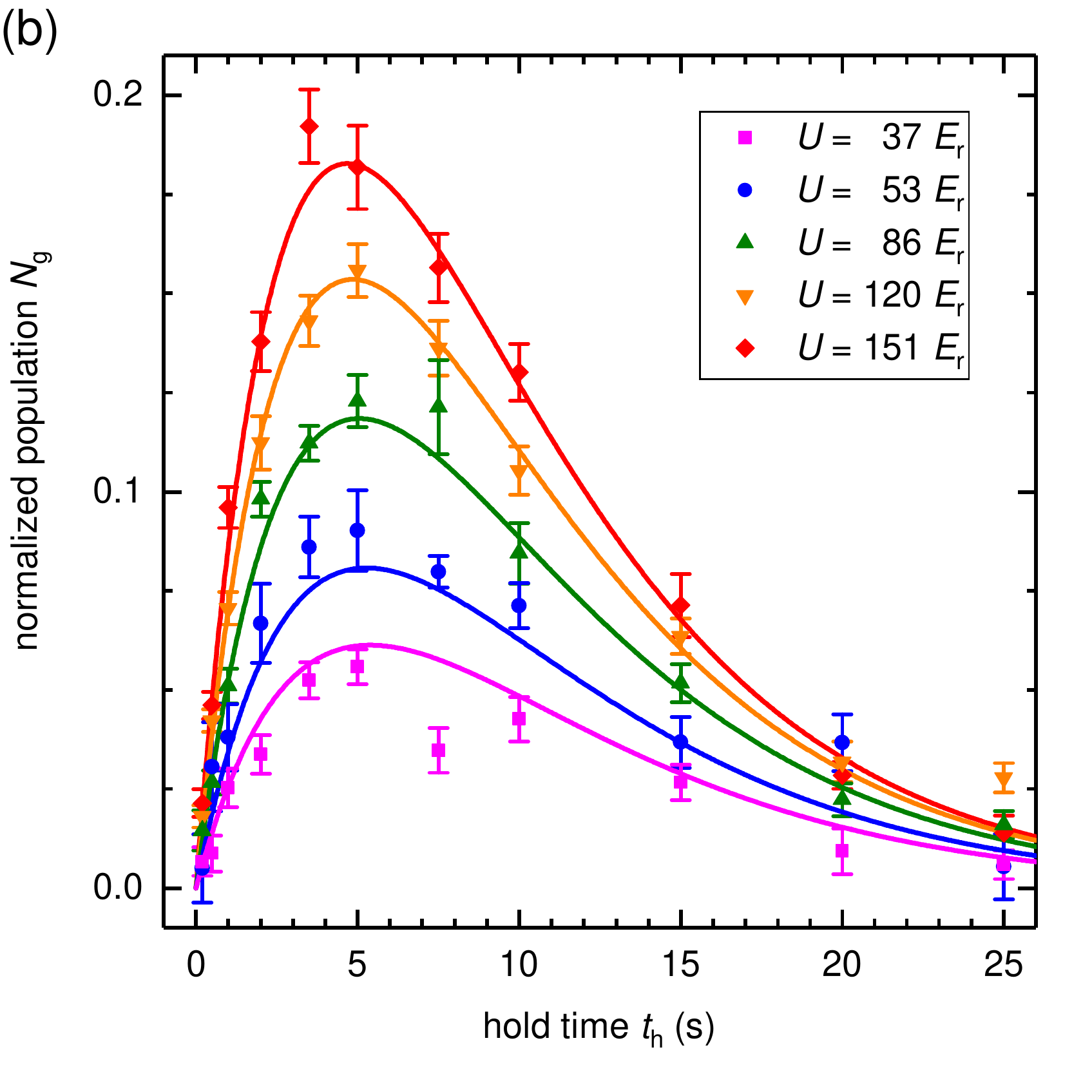}}
	\caption{(color online)
		Measured populations (a) in the metastable states $\state{3}{P}{0,2}$ and (b) in the ground state $\state{1}{S}{0}$ for samples prepared in the excited state at different effective lattice depths $\Utherm$ as a function of hold time $\thold$.
		Results of a combined fit of a rate equation model (see text for details) are shown as colored lines.
		The inset to (a) shows the fraction of atoms in the ground state as derived from this model.
		For reference, the solid black line shows the decay of ground-state samples (see Fig.~\ref{fig:measurement_gs}).}
	\label{fig:measurement_es}
\end{figure*}
Before studying decay from the excited state, we first investigate the loss of ground-state atoms from the trap.
Figure~\ref{fig:measurement_gs} shows the normalized ground-state population $\Ngnd$ as a function of hold time for different lattice depths.
We observe an exponential decay, which is consistent with density-independent losses, and no discernible influence of lattice depth.
Fits allowing for a density-dependent loss rate showed no consistent results for the loss coefficient and were, therefore, not pursued further.
The trap lifetime is about \SI{6}{\second}, which is probably limited by the residual pressure in our vacuum system.
Owing to the decreasing signal-to-noise ratio and technical restrictions, we limit the hold time to $\thold \leq \SI{25}{\second}$ throughout our measurements.

When the samples are prepared in the excited state as described above, population dynamics as shown in Fig.~\ref{fig:measurement_es} are observed instead.
We clearly observe decay $\state{3}{P}{0} \rightarrow \state{1}{S}{0}$, as, in addition to enhanced losses in the excited state, atoms emerge in the ground state.
The population in the ground state becomes substantial after several seconds, especially for deep lattice potentials. 

For a quantitative analysis of our measurements, we model the time evolution of the two populations by a pair of coupled rate equations
\begin{subequations}
	\begin{eqnarray}
	\dot{\Nexc} & = & - \traplossES \Nexc - \left( \decayrateOTHER + \decayratecoeffLAT \Utherm \right) \Nexc \label{eq:rate_equation_es} \\
	\dot{\Ngnd} & = & - \traplossGS \Ngnd + \left( \decayrateOTHER + \decayratecoeffLAT \Utherm \right) \Nexc \label{eq:rate_equation_gs}
	\end{eqnarray}
	\label{eq:rate_equation}
\end{subequations}
that describe lattice-independent losses from the trap at different rates, $\traplossGS$ and $\traplossES$, as well as decay from the excited state to the ground state.
The rate of the latter is the sum of a lattice-independent contribution $\decayrateOTHER$ and a lattice-induced contribution $\decayratecoeffLAT \Utherm$.
The analytic solution to these rate equations reads
\begin{subequations}
	\begin{eqnarray}
	\Nexc(t) & = & \Nexc(0)
	               \exp\left( -\left[ \traplossES + \decayrateOTHER + \decayratecoeffLAT \Utherm \right] t \right) \\
	\Ngnd(t) & = & \left( \Nexc(0)
	               \frac{ 1 - \exp\left( - \left[ \traplossDIFF + \decayrateOTHER + \decayratecoeffLAT \Utherm \right] t \right) }
	               { 1 + \traplossDIFF / \left( \decayrateOTHER + \decayratecoeffLAT \Utherm \right) } \right.\nonumber\\
	         &   & \left. + \Ngnd(0) \vphantom{\frac{\traplossDIFF}{\traplossDIFF}} \right) \exp\left( -\traplossGS t \right) 
	\end{eqnarray}
	\label{eq:model_solution}
\end{subequations}
where $\traplossDIFF = \traplossES - \traplossGS$.
We perform a nonlinear least-squares fitting of Eqs.~(\ref{eq:model_solution}) to the entire data set as shown in Figs.~\ref{fig:measurement_gs} and \ref{fig:measurement_es}.
The initial populations $\Nany(0)$ are set to $N_0$ if the state is occupied at $\thold = 0$ or to zero otherwise.
The uncertainties of the parameters given in the remainder of this section result from this fit.
We have checked for potential lattice-induced losses from the trap by adding a contribution $\tilde{\gamma}\Utherm$ to the loss rates in Eqs.~(\ref{eq:rate_equation_es}) and (\ref{eq:rate_equation_gs}), but find no indication of such an effect.

As seen most clearly in Fig.~\ref{fig:measurement_es}(b), the rate equations (\ref{eq:rate_equation}) describe the measured population dynamics and, in particular, the dependence of the rate of decay $\state{3}{P}{0} \rightarrow \state{1}{S}{0}$ on lattice depth very well.
Our findings indicate that lattice-induced decay is the dominant decay mechanism at typical lattice depths.

We find $\traplossGS = \SI{0.1650(14)}{\per\second}$ and $N_0 = \SI{1.204(5)}{}$ for our specific apparatus and experimental procedure.
Trap losses in the excited state are slightly larger than in the ground state [$\traplossES/\traplossGS = \SI{1.101(13)}{}$]; discussion of this difference is beyond the scope of this work, however.
It is straightforward to increase the lifetime of atoms in the lattice by reducing background pressure to facilitate spectroscopy at interrogation times of several seconds or more.

As key results, we determine the decay rate $\state{3}{P}{0} \rightarrow \state{1}{S}{0}$ due to inelastic scattering of lattice laser radiation and the rate stemming from other sources including spontaneous decay (see Sec.~\ref{sec:lifetime}):
\begin{eqnarray*}
	\decayratecoeffLAT & = & \SI{556(15)d-6}{\per\second}\,\Erec^{-1}
	\label{eq:res_scatterrate} \\
	\decayrateOTHER    & = & \SI{5.2(12)d-3}{\per\second}
	\label{eq:res_decayrate}
\end{eqnarray*}
As a practical note, the lattice-induced rate equals $\decayrateOTHER$ at a lattice depth $\Utherm \approx 9.4\,\Erec$.
\section{\label{sec:scat}Calculation of scattering rates}
In addition to our experimental results, we calculate the off-resonant scattering rates of lattice laser radiation from atomic data in order to investigate the other processes shown in Fig.~\ref{fig:scattering_summary}, to which our experiment is not sensitive.
The scattering rate $\state{3}{P}{0} \rightarrow \state{3}{P}{1}$ measured in the previous section is used to verify the results of our calculation.
\subsection{\label{subsec:scat_theory}Theory}
The rate $\Gamma_{i \rightarrow f}$ of an atom in an initial state $i$ off-resonantly scattering linearly polarized incoming radiation with intensity $I$ and being transferred to a final state $f$ is given by the Kramers--Heisenberg formula \cite{bra03b}
\begin{equation}
	\Gamma_{i \rightarrow f} = \frac{I {\omega^{\prime}}^3}{\left(4\pi \epsilon_{0} \right)^2 c^4 \hbar^3} \frac{8\pi}{3} \sum_{q=-1}^{1} \left| D_q^{(i \rightarrow f)} \right|^2
	\label{eq:sc_rate}
\end{equation}
with
\begin{equation}
	\begin{split}
		D_q^{(i \rightarrow f)} = \sum_{k} &\left( \matrixelement{f}{d_{q}}{k} \frac{\matrixelement{k}{d_{0}}{i}} {\omega_{ki} - \omega} + \right. \\
		                                   &\left. \matrixelement{f}{d_{0}}{k} \frac{\matrixelement{k}{d_{q}}{i}} {\omega_{ki} + \omega^{\prime}}  \right)
	\end{split}
	\label{eq:sc_amplitude}
\end{equation}
where $d_q = -e r_q$ are the elements of the electric dipole operator in spherical tensor notation, $\omega$ and $\omega^{\prime}$ are the angular frequencies of the incoming and scattered radiation, respectively, and $q$ is the polarization state of the scattered radiation in spherical tensor notation, where the polarization axis of the incoming light is used as quantization axis.
The sum is over all intermediate states $k$, and $\omega_{ki}$ is the frequency of the transition $i \rightarrow k$.

The specific case of linearly polarized lattice light is realized in our lattice clock as well as in many others.
Typically, a bias magnetic field is applied parallel to the polarization vector of the lattice during spectroscopy.

The two terms in Eq.~(\ref{eq:sc_amplitude}) can be interpreted physically as absorbing an incoming photon before emitting a scattered photon or vice versa.
Note that Eqs.~(\ref{eq:sc_rate}) and (\ref{eq:sc_amplitude}) neglect contributions from multiphoton transitions and non--electric dipole transitions.

The dependence of Eq.~(\ref{eq:sc_amplitude}) on the magnetic substates can be separated using the Wigner--Eckart theorem \cite{edm96}:
\begin{equation} 
	\begin{split}
		\matrixelement{k^{\prime}, F^{\prime} m_F^{\prime}}{d_{q}}{k,  F m_F} = 
		    (-1)^{F^{\prime} - m_F^{\prime}}
		\\ \times
		    \threejsymbol{F^{\prime}}{1}{F}{-m_F^{\prime}}{q}{m_F}
		    \reducedmatrixelement{k^{\prime}, F^{\prime}}{d}{k, F}
	\end{split}
	\label{eq:wigner_eckart}
\end{equation}
where the term denoted by round brackets is the Wigner $3j$-symbol. $\reducedmatrixelement{k^{\prime}, F^{\prime}}{d}{k, F}$ is the reduced matrix element with respect to total angular momentum $\boldsymbol{F}$.

As the electric dipole operator $d$ acts on the subspace of orbital angular momentum $\boldsymbol{L}$, the reduced matrix element can be expressed for the decoupled angular momenta using the well-known relation \cite{edm96}
\begin{equation}
	\begin{split}
		\reducedmatrixelement{k^{\prime}, (j_1^{\prime} j_2) j_3^{\prime}}{d}{k, (j_1 j_2) j_3} = 
		    (-1)^{j_1^{\prime} + j_2 + j_3 + 1}
		\\ \times
		    \sqrt{ \left( 2 j_3^{\prime} + 1 \right) \left( 2 j_3 + 1 \right)}
		    \sixjsymbol{j_1^{\prime}}{j_3^{\prime}}{j_2}{j_3}{j_1}{1}
		    \reducedmatrixelement{k^{\prime}, j_1^{\prime}}{d}{k, j_1}
	\end{split}
	\label{eq:wigner_6j}
\end{equation}
if the operator $d$ acts only on part $1$ of the system, where $\boldsymbol{j}_3 = \boldsymbol{j}_1 + \boldsymbol{j}_2$, and the curly brackets denote the Wigner $6j$-symbol.

The reduced matrix element $\reducedmatrixelement{k}{d}{i}$ is related to the Einstein coefficient $A_{ki}$ for decay $k \rightarrow i$ by \cite{hil02a}
\begin{equation}
	\left|\reducedmatrixelement{k}{d}{i}\right|^2 = \frac{3 \epsilon_0 h c^3}{2\omega_{ki}^3} g_k A_{ki}
	\label{eq:einstein_coefficient}
\end{equation}
where $g_k$ is the degeneracy factor of the state $k$.

We calculate the scattering rates according to Eq.~(\ref{eq:sc_rate}) using the line strengths reported in Ref.~\cite{nic15} for the $(4d5s)\,\state{3}{D}{}$ states, in Ref.~\cite{san10a} for the state $(5s6s)\,\state{3}{S}{1}$, and listed in the supplement to Ref.~\cite{mid12a} otherwise.
Similar to our previous publication \cite{mid12a} we adjust the line strengths of the $\state{3}{D}{}$ continuum and of the $(5s5d)\,\state{3}{D}{}$ states to reproduce well-known parameters, specifically the differential dc-Stark polarizability and the magic wavelength.
For scattering processes $\state{3}{P}{0} \rightarrow \state{3}{P}{J}$ ($J\neq0$), the relative signs of the products of reduced matrix elements in Eq.~(\ref{eq:sc_amplitude}) can be determined from Eq.~(\ref{eq:wigner_6j}).
We focus solely on the case $m_F = 9/2$ in the following, because the scattering processes from $m_F = -9/2$ are equivalent for typical bias magnetic fields.

\subsection{\label{subsec:scat_polarizability}Dynamic polarizability and lattice depth}
The scattering amplitudes in Eq.~(\ref{eq:sc_amplitude}) are closely related to the dynamic polarizability of the state $i$, which is given by \cite{bra03b}
\begin{equation}
	\alpha_i = \frac{2}{\hbar} \sum_{k} \frac{\omega_{ki} \left| \matrixelement{k}{d_{0}}{i} \right|^{2}} {\omega_{ki}^{2} - \omega^{2}}
	\label{eq:dyn_polarizability}
\end{equation}
for linearly polarized light.

Note in particular that the Rayleigh scattering amplitudes are related to this polarizability by
\begin{equation}
	D_0^{(i \rightarrow i)} = \hbar \alpha_{i}
	\label{eq:rayleigh_scattering_and_polarizability}
\end{equation}
if photon recoil is neglected ($\omega^{\prime} = \omega$).
As a consequence, the elastic scattering amplitudes of the two states $\state{1}{S}{0}$ and $\state{3}{P}{0}$ used in lattice clocks are necessarily equal at any magic wavelength, where, by definition, the difference of the dynamic polarizabilities vanishes.

Furthermore, we use Eq.~(\ref{eq:rayleigh_scattering_and_polarizability}) to infer the optical potential depth for a given intensity of the light field from the elastic scattering rates.
The optical dipole potential depth at an intensity $I$ of the light field is given by
\begin{equation}
	U = \frac{1}{2\epsilon_{0}c} \alpha_i I\mbox{.}
	\label{eq:dipole_potential}
\end{equation}
\subsection{\label{subsec:scat_results}Results}
The rates of all Rayleigh ($i = f$) and Raman ($i \neq f$) scattering processes of laser radiation at the magic wavelength near \SI{813}{\nano\meter} that occur for $\Sr$ atoms (see Fig.~\ref{fig:scattering_summary}) are summarized in Table~\ref{tab:all_rates}.
\begin{table}[tb]
	\caption{Calculated off-resonant scattering rates of lattice laser radiation by an atom in the magnetic substate $m_F = 9/2$ of the state $i$.
		The final state of the the atom is denoted by $f$ with hyperfine and magnetic quantum numbers $F^{\prime}$ and $m_F^{\prime}$, respectively.
		Rates are given for an intensity corresponding to an optical dipole potential of $\Utherm = 1\,\Erec$.}
	\label{tab:all_rates}
	\begin{ruledtabular}
		\begin{tabular}{cccc}
			$i \rightarrow f$							& $F^\prime$	& $m_F^\prime$	& $ \Gamma / (\SI{d-4}{\per\second}$)	\\\hline\\
			$\state{1}{S}{0}\rightarrow\state{1}{S}{0}$	& $ 9/2$		& $ 7/2$		& \SI{3e-16}{}							\\
														&				& $ 9/2$		& \SI{5.57}{}							\\\\
			$\state{3}{P}{0}\rightarrow\state{3}{P}{0}$	& $ 9/2$		& $ 7/2$		& \SI{5e-10}{}							\\
														&				& $ 9/2$		& \SI{5.57}{}							\\\\
			$\state{3}{P}{0}\rightarrow\state{3}{P}{1}$	& $ 7/2$		& $ 7/2$		& \SI{1.99}{}							\\
														& $ 9/2$		& $ 7/2$		& \SI{0.45}{}							\\
														&				& $ 9/2$		& \SI{1e-10}{}							\\
														& $11/2$		& $ 7/2$		& \SI{0.05}{}							\\
														&				& $ 9/2$		& \SI{6e-10}{}							\\
														&				& $11/2$		& \SI{2.49}{}							\\\\
			$\state{3}{P}{0}\rightarrow\state{3}{P}{2}$	& $ 7/2$		& $ 7/2$		& \SI{0.37}{}							\\
														& $ 9/2$		& $ 7/2$		& \SI{0.26}{}							\\
														&				& $ 9/2$		& \SI{0.76}{}							\\
														& $11/2$		& $ 7/2$		& \SI{0.08}{}							\\
														&				& $ 9/2$		& \SI{0.53}{}							\\
														&				& $11/2$		& \SI{0.50}{}							\\
														& $13/2$		& $ 7/2$		& \SI{0.01}{}							\\
														&				& $ 9/2$		& \SI{0.11}{}							\\
														&				& $11/2$		& \SI{0.22}{}							\\
		\end{tabular}
	\end{ruledtabular}
\end{table}

Concerning Rayleigh scattering, we find equal rates of $\SI{5.57d-4}{\per\second}\cdot(\Utherm/\Erec)$ in both states, $\state{1}{S}{0}$ and $\state{3}{P}{0}$, as expected due to Eq.~(\ref{eq:rayleigh_scattering_and_polarizability}).
We infer a dynamic polarizability $\alpha = \SI{4.622d-39}{\joule\volt^{-2}\meter^2}$ (or $\SI{280.4}{}$~a.u.) of these states using the same equation.
These results are in excellent agreement with previous publications \cite{mar13a, shi15}.

The total rate of Raman scattering $\state{3}{P}{0} \rightarrow \state{3}{P}{1}$ is $\SI{4.98d-4}{\per\second} \cdot (\Utherm/\Erec)$, which is similar in magnitude to the Rayleigh scattering rate.
This value agrees well with our experimental observations (see Sec.~\ref{sec:exp}) as well as previous predictions \cite{mar13a}.
Moreover, our calculation yields the individual scattering rates to the different hyperfine states and magnetic substates, which are listed in Table~\ref{tab:all_rates}.

\begin{figure}[tb]
	\centerline{\includegraphics{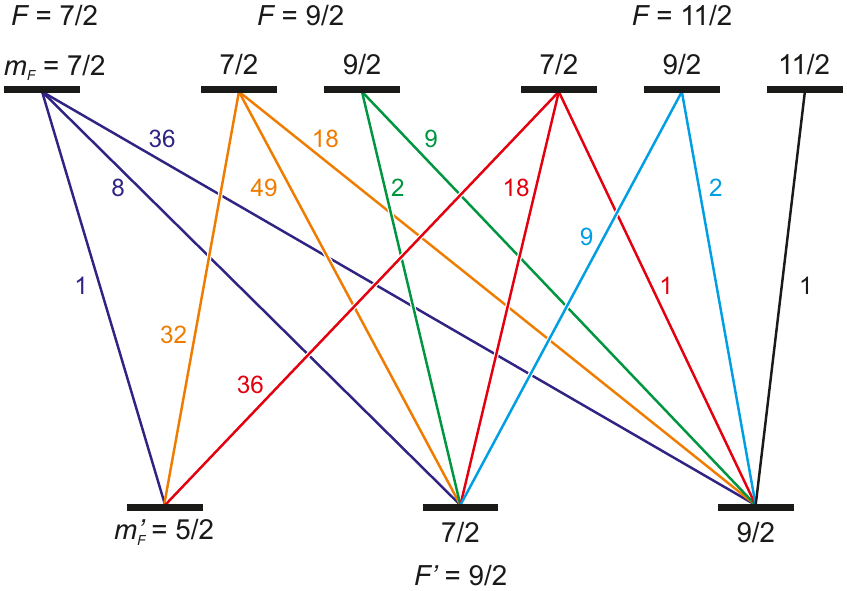}}
	\caption{(color online)
		Branching ratios of the decay $\state{3}{P}{1} \rightarrow \state{1}{S}{0}$ for all hyperfine states and magnetic sublevels that are accessible from the magnetic sublevel $m_F=9/2$ of the state $\state{3}{P}{0}$ by Raman scattering of linearly polarized lattice laser light.}
	\label{fig:fig_branching_ratios_3P1}
\end{figure}

The effective rates of lattice-induced decay to the three accessible magnetic substates of the ground state, $m_F = 5/2$, $7/2$, and $9/2$, are listed in Table~\ref{tab:decay_rates}.
We determine them by combining the rates in Table~\ref{tab:all_rates} with the branching ratios of the subsequent radiative decay to the ground state, which are shown in Fig.~\ref{fig:fig_branching_ratios_3P1} and result from Eqs.~(\ref{eq:wigner_eckart}) through (\ref{eq:einstein_coefficient}). 
We find that lattice-induced decay to the ground state mainly populates the original magnetic substate, $m_F = 9/2$.
\begin{table}
	\caption{\label{tab:decay_rates}Calculated lattice-induced decay rates $\decayrateLAT$ from the magnetic substate $m_F=9/2$ of the excited state to the accessible magnetic substates of the ground state, with magnetic quantum numbers $m_F^{\prime}$.
		Rates are given for an intensity corresponding to an optical dipole potential $\Utherm = 1\,\Erec$.}
	\begin{ruledtabular}
		\begin{tabular}{cccc}
			$m_F^\prime$											& $5/2$			& $7/2$			& $9/2$			\\
			$\decayrateLAT(m_F^\prime) / (\SI{d-4}{\per\second})$	& \SI{0.22}{}	& \SI{0.59}{}	& \SI{4.17}{}	\\
		\end{tabular}
	\end{ruledtabular}
\end{table}

For Raman scattering $\state{3}{P}{0} \rightarrow \state{3}{P}{2}$, we predict a total rate of $\SI{2.84d-4}{\per\second} \cdot (\Utherm/\Erec)$, which is about a factor of two smaller than for Rayleigh scattering.
Again, this value is in agreement with previous predictions \cite{mar13a}.
Table~\ref{tab:all_rates} lists the individual scattering rates to the various accessible hyperfine states and different magnetic substates of the final state.

Lastly, Raman scattering into the magnetic substate $m_F=7/2$ of the original state, $\state{1}{S}{0}$ or $\state{3}{P}{0}$, is strongly suppressed by destructive interference of the different paths  (see Table~\ref{tab:all_rates}).
Raman scattering to final states in the $(5s4d)\,\state{3}{D}{}$ manifold has not been considered, because it is not possible via single-photon electric-dipole transitions.
Therefore, only Rayleigh scattering and Raman scattering $\state{3}{P}{0} \rightarrow \state{3}{P}{1,2}$ are relevant in $\Sr$ lattice clocks.
\section{\label{sec:lifetime}Radiative lifetime of the excited state}
The observed lattice intensity--independent rate $\decayrateOTHER$ of excited-state decay (see Sec.~\ref{subsec:exp_results}) stems from the natural lifetime of the excited state and other lattice-independent processes.
Since care is taken to block out any laser radiation, apart from the lattice laser beam and the interrogation laser beam driving the reference transition, while the atoms are held in the trap, optical pumping on the transition $\state{3}{P}{0} \textendash \state{3}{D}{1}$ by BBR followed by spontaneous decay to the ground state via the state $\state{3}{P}{1}$ is the most likely process to compete with direct spontaneous decay of the excited state.

The quenching rate $\decayrateBBR$ due to BBR can be estimated \cite{yas04} from the BBR-induced rate of excitation and the branching ratio $R_1$ of spontaneous decay to the state $\state{3}{P}{1}$.
The former can be described by the lifetime $\tau(\state{3}{D}{1}) = \SI{2.18(1)d-6}{\second}$ \cite{nic15} of the state $\state{3}{D}{1}$, its degeneracy $2J + 1$, and the branching ratio $R_0$ of spontaneous decay to the state $\state{3}{P}{0}$, assuming Russell--Saunders coupling.
Since the frequencies of the transitions $\state{3}{D}{1} \textendash \state{3}{P}{J}$ vary significantly due to fine-structure splitting of the multiplet \cite{san10a}, the cubic frequency dependence of the spontaneous emission rates must be taken into account when determining any of the branching ratios $\branchingratio_J$ using Eq.~(\ref{eq:wigner_6j}).
This results in branching ratios of \SI{59.65}{\percent}, \SI{38.52}{\percent}, and \SI{1.82}{\percent} to $J=0$, $1$, and $2$, respectively.
We thus find a quenching rate
\begin{equation}
	\decayrateBBR(T) = \frac{3 \branchingratio_0 \branchingratio_1}{\tau(\state{3}{D}{1})} \frac{1}{\exp\left(\hbar\omega / \kB T\right) - 1}
	\label{eq:BBR_quenchrate}
\end{equation}
at a temperature $T$, where $\omega$ is the transition frequency $\state{3}{P}{0} \textendash \state{3}{D}{1}$.
This results in $\decayrateBBR = \SI{2.23(14)d-3}{\per\second}$ at the temperature $T=\SI{294.5(10)}{\kelvin}$ of our experimental apparatus.

We attribute the remaining rate 
\begin{equation*}
	\decayrateSPO = \decayrateOTHER - \decayrateBBR = \SI{3.0(13)d-3}{\per\second}
\end{equation*}
to spontaneous decay of the state $\state{3}{P}{0}$.
The corresponding lifetime $\tau(\state{3}{P}{0})= \SI{330(140)}{\second}$ is marginally in agreement with the value of $\SI{145(40)}{\second}$ predicted in Ref.~\cite{boy07a}.
To the best of our knowledge, this is the first direct experimental measurement of the lifetime of this state in $\Sr$.
\section{\label{sec:consequences}Effects on lattice clocks}
The relevant off-resonant scattering processes of laser radiation at the magic wavelength near $\SI{813}{\nano\meter}$, which have been identified in Sec.~\ref{sec:scat}, affect lattice clocks in several different ways.

Although Rayleigh scattering is elastic, it may still lead to damping of the coherence of the atomic superposition state.
It has been shown previously that the rate of coherence damping depends on the difference of the elastic scattering amplitudes \cite{uys10}, and the situation in optical lattice clocks has been discussed in Ref.~\cite{mar13a}.
We point out that the difference of the Rayleigh scattering amplitudes between the two states $\state{1}{S}{0}$ and $\state{3}{P}{0}$ is proportional to their differential dynamic polarizability and vanishes at the magic wavelengths (see Sec.~\ref{subsec:scat_polarizability}).
Therefore, Rayleigh scattering of lattice laser radiation does not cause significant decoherence in optical lattice clocks.

The two Raman scattering processes result in decoherence of the atomic superposition state, as well as depopulation of the excited state $\state{3}{P}{0}$.
They do not cause systematic frequency shifts directly, as they are not sensitive to the phase of the superposition state.
However, the maximum slope of the spectroscopic signal is reduced.
In Ramsey spectroscopy, these effects manifest as a reduction of fringe contrast if the duration of an excitation pulse is small compared to the inverse scattering rate.
For excitation with a single, long pulse, the line shape is modified more intricately; it is broadened, but remains symmetric with respect to the detuning of the interrogation laser from resonance.

Moreover, atoms transferred to the ground state and, possibly, to the metastable state $\state{3}{P}{2}$ are still registered during state detection.
Owing to the stochastic nature of the scattering process, this typically results in intrinsic noise of the detected atomic populations and thus reduces the signal-to-noise ratio.
For the metastable state $\state{3}{P}{2}$, this can be avoided by using a more sophisticated detection scheme to discern atoms in the two states $\state{3}{P}{0}$ and $\state{3}{P}{2}$, e.g., by selective repumping on the transition $\state{3}{P}{0} \textendash \state{3}{D}{1}$ as demonstrated in ytterbium lattice clocks \cite{bar07a, lem09}.

The populations created, through Raman scattering, in the states $\state{3}{P}{2}$ and $\state{1}{S}{0}$ themselves may also disturb an optical lattice clock.
In particular, atoms having decayed to the ground state are highly susceptible to the interrogation laser, whereas the metastable state $\state{3}{P}{2}$ can effectively be considered a dark state in this respect.

Our calculations in Sec.~\ref{sec:scat} have shown that lattice-induced decay to the ground state most likely returns an atom to the magnetic sublevel $m_F = \pm9/2$ (see Table~\ref{tab:decay_rates}).
The consequences on spectroscopy are quite similar to those of decoherence and depopulation of the superposition state.
For Ramsey spectroscopy, the atoms having decayed during the free evolution time are resonantly excited by the final $\pi/2$-pulse, resulting in reduced fringe contrast.
In case of Rabi spectroscopy with long excitation pulses, coherently driving the reference transition in those atoms after decay modifies the line shape further.
Likewise, this degrades the slope of the error signal and, in some cases, the observable line width, but does not give rise to systematic frequency shifts.

In contrast, population of the other magnetic substates ($m_F\neq\pm9/2$) in the ground state does cause line pulling if the laser detuning is varied to derive an error signal.
It can be reduced by choosing the spectral bandwidth of the excitation pulse or pulses either much smaller or much larger than the frequency detuning of the transitions from adjacent magnetic substates due to the bias magnetic field.
Moreover, $\Sr$ lattice clocks typically monitor the transition frequencies from both stretched magnetic substates \cite{tak06}; center frequency and frequency splitting are used to stabilize the frequency of the interrogation laser and monitor the bias magnetic field, respectively.
For reasons of symmetry, line pulling due to populating different magnetic substates by Raman scattering mainly affects the frequency splitting rather than the center frequency.
The linear Zeeman splitting is also used in correcting for the quadratic Zeeman effect in our clock \cite{fal11}, but the effect of Raman scattering on this correction of the transition frequency is well below one part in $10^{18}$ at typical bias field strengths.
Line pulling can be avoided entirely for Ramsey spectroscopy by varying the relative phase of the final pulse \cite{ram51,let04} rather than the laser detuning to generate an error signal.

In either case, atoms excited by the interrogation laser after Raman scattering give rise to quantum projection noise, especially for Ramsey spectroscopy.

Aside from the scattering processes shown in Fig.~\ref{fig:scattering_summary}, we do not investigate changes of the motional state of an atom in the optical lattice by off-resonant scattering in detail in this work.
They may lead to line pulling and systematic frequency shifts due to the increased tunneling in excited axial motional states.
However, their relative strength is small in the Lamb--Dicke regime. They are not expected to cause substantial decoherence themselves, as the external potential is nearly identical for the two clock states near the magic wavelength.

Last but not least, atoms that are incoherently transferred to different states by Raman scattering may cause systematic shifts of the transition frequency by disturbing the remaining atoms.
For instance, interactions with atoms in the original superposition state at the same lattice site are generally no longer suppressed by the Pauli principle and lead to collision-induced systematic frequency shifts at high atom density.
\section{\label{sec:discussion}Discussion}
Systematic frequency shifts of the reference transition due to off-resonant photon scattering from the lattice need to be evaluated under actual operating conditions, but they are not expected to become a fundamental problem at long interrogation times.
As discussed in the previous section, such shifts may be caused by the scattering atoms indirectly, e.g., by atomic interactions, or by differences of the scattering rates between the magnetic substates $m_F = \pm9/2$, which are small at typical bias magnetic fields for linear polarization of the lattice.
On-site atomic interactions can be avoided by using a three-dimensional lattice to suppress double occupancy \cite{cam17} or by operating in a low-density regime as in our clock.

Collapse of the atomic superposition state as well as quantum projection noise arising from atoms excited by the interrogation laser after decay to the ground state reduce the signal-to-noise ratio substantially at long interrogation times or in deep lattice potentials.
Our investigations have shown that off-resonant photon scattering from the lattice gives rise to time constants of few $\SI{10}{\second}$ at typical lattice depths of several $10\,\Erec$, which well exceeds contributions from other sources, such as spontaneous decay or pumping by BBR.
This ultimately degrades the minimum frequency instability and may prevent lattice clocks from reaching the QPN limit given by Eq.~(\ref{eq:QPNL}) and from exploiting squeezing, or entanglement in general, to achieve frequency instability below that limit under these conditions.
Therefore, the decreasing signal-to-noise ratio is the most important problem for the operation of lattice clocks at interrogation times of several seconds and beyond.

Operating a clock with a shallow lattice potential obviously reduces the scattering rates, but increases other systematic effects such as frequency shifts due to tunneling.
However, it has been shown previously that gravity can be exploited in vertically oriented lattice geometries to greatly reduce tunneling by lifting the degeneracy of adjacent lattice sites, which allows control of the resulting frequency shifts to below $\SI{1}{\milli\hertz}$ at a potential depth of only $5\,\Erec$ for strontium lattice clocks \cite{lem05}.
At this lattice depth, the induced photon scattering rates are comparable to the natural decay rate of the excited state, with time constants of a few $\SI{100}{\second}$ (see Secs.~\ref{sec:exp} and \ref{sec:lifetime}).

Lattice clocks in microgravity environments or using multidimensional lattices require more intense lattice light fields and thus suffer from increased off-resonant scattering rates.
For these systems, in particular, further investigations into controlling motional effects in lattice clocks or alternative solutions are required.

Off-resonant scattering rates are generally different if the optical lattice is operated at a magic wavelength other than the widely used one near $\SI{813}{\nano\meter}$, which is studied here.
The relative strengths of the scattering processes can be quite different \cite{mar13a}, although the dependence of Eq.~(\ref{eq:sc_rate}) and the recoil energy $\Erec$ on laser wavelength is generally not favorable to shorter wavelengths. 

However, the light intensity experienced by the atoms can be greatly reduced by operating the lattice at a blue-detuned magic wavelength \cite{ger10b, pic10}, e.g., near $\SI{390}{\nano\meter}$ \cite{tak09}, where the atoms are trapped in the nodes of the light field.
This provides a highly interesting option to avoid off-resonant scattering of lattice laser radiation and merits further study.
\section{\label{sec:conclusion}Conclusion}
Off-resonant photon scattering from the lattice becomes relevant if optical lattice clocks, with typical lattice depths of several $10\,\Erec$, are  operated at interrogation times of several seconds or more.
Relevant scattering processes near the magic wavelength of $\SI{813}{\nano\meter}$ are Rayleigh scattering and Raman scattering $\state{3}{P}{0} \rightarrow \state{3}{P}{1,2}$.
The former Raman scattering process, $\state{3}{P}{0} \rightarrow \state{3}{P}{1}$, gives rise to lattice-induced decay to the ground state, which we have observed experimentally.
These observations and complementary predictions based on atomic data show that all of these processes occur at rates of several $\SI{d-2}{\per\second}$ at typical lattice depths, exceeding the rates of natural and BBR-induced decay of the excited state.
Furthermore, our measurements yield an experimental value $\tau = \SI{330(140)}{\second}$ for the natural lifetime of the excited state $\state{3}{P}{0}$, which is in marginal agreement with predictions \cite{boy07a}.
To the best of our knowledge, this is the first experimental measurement of the natural lifetime of this state.

In optical lattice clocks, Rayleigh scattering does not cause significant decoherence of the atomic superposition state used for spectroscopy, because the optical lattice is operated near a magic wavelength.
However, collapse of the atomic superposition state and transfer of population to different internal states due to Raman scattering eventually limit the achievable frequency instability and thus useful interrogation times in optical lattice clocks at the presently used lattice depths.
Lattice-induced decay to the ground state, in particular, modifies the observed line shape and contributes to quantum projection noise.
Raman scattering also counteracts suppression of atomic interactions and may thus lead to systematic frequency shifts.
We have pointed out several ways to overcome these problems.
Most importantly, vertical lattice geometries allow operating clocks with $10^{-18}$ fractional accuracy at lattice depths as shallow as $5\,\Erec$ \cite{lem05}, which reduces the photon scattering rates to levels comparable to those of natural decay or BBR-induced pumping.

One may wonder whether a limit on the interrogation time set by the scattering of lattice photons, which may be expected at around \SI{100}{\second} in a strontium lattice clock, will give ion clocks an intrinsic advantage over lattice clocks once interrogation lasers achieve comparable coherence times.
This can only be the case for ions that have much narrower transitions, such as the $\mathrm{Yb}^+$ octupole transition with an excited-state lifetime of several years \cite{rob97, bie98} or $\mathrm{B}^+$ \cite{lud15}, but not for systems like the $\mathrm{Al}^+$ ion with an excited-state lifetime of only several tens of seconds \cite{ros07}.
In ion clocks, there are other limitations of the interrogation time, including heating in the trap, which must be overcome.
Longer interrogation times $T_\mathrm{i}$ increase not only the observed line quality factor $Q \propto T_\mathrm{i}$ (up to its natural value) but also the measurement cycle duration $T_\mathrm{c} \approx T_\mathrm{i}$, thus leading to a frequency instability $\sigma_y \propto T_\mathrm{i}^{-1/2}$ according to Eq.~(\ref{eq:QPNL}). However, frequency instability scales with atom number in the same way, $\sigma_y \propto N^{-1/2}$.
Lattice clocks easily exceed ion clocks by factors of more than 100 in the number of atoms being interrogated at the same time.
To achieve comparable frequency instability, the interrogation time in an ion clock must be extended by that same factor.
This would require interrogation times of several hours or more, which seems to be well beyond reach in the foreseeable future.
Therefore, we think that, although off-resonant scattering of lattice laser radiation cannot be avoided in optical lattice clocks and may cause limitations, it is unlikely to pose a strong disadvantage compared to single-ion clocks.
\begin{acknowledgments}
This work is supported by QUEST, by DFG within CRC 1227 (DQ-mat, project B02) and RTG 1729. The results in this contribution partly come from the project EMPIR 15SIB03 OC18. This project has received funding from the EMPIR programme co-financed by the Participating States and from the European Union's Horizon 2020 research and innovation programme.
\end{acknowledgments}

\end{document}